\documentclass[aps,amsmath,onecolumn,prl,showpacs]{revtex4}

\usepackage[]{graphicx}
\usepackage{amsmath}
\begin{document}

% \DOIsuffix{theDOIsuffix}
% \Volume{XX}
% \Issue{1}
% \Month{01}
% \Year{2003}
% \pagespan{3}{}
% \Receiveddate{15 November 2003}
% \Reviseddate{30 November 2003}
% \Accepteddate{2 December 2003}
% \Dateposted{3 December 2003}
%\keywords{Microcavities, excitons, polaritons, Rabi splitting, Bose-Einstein condensation, parametric scattering.}

\title{Fifteen years of microcavity polaritons}

\email[]{vincenzo.savona@epfl.ch}
\author{Vincenzo Savona}
\affiliation{Institute of Theoretical Physics, Ecole Polytechnique F\'ed\'erale de Lausanne EPFL, CH-1015 Lausanne, Switzerland}

\date{\today}

\begin{abstract}
We present an overview of the advances in the physics of polaritons in semiconductor microcavities, starting from their first discovery in 1992. After summarizing the research during the early years, when the basic optical properties were investigated, we describe the important results related to the phenomenon of parametric polariton scattering, highlighting its link to non-classical polariton physics. We conclude with a perspective view of the future research in the domain, focusing on the exciting developments in the direction of polariton quantum collective phenomena and quantum nanodevices.
\end{abstract}


\maketitle                   % Produces the title.

\tableofcontents

\section{Introduction}

In semiconductors with a direct interband optical transition, the fundamental electronic excitations above the ground state are excitons, namely hydrogen-like bound electron-hole pairs. In a perfect bulk semiconductor material, however, a correct description of the excited states must include the linear coupling of excitons to the electromagnetic field. This coupling gives rise to normal modes that are linear superpositions of one exciton and one photon mode, called {\em exciton-polaritons}. Exciton-polaritons are the actual excited states of a bulk semiconductor. 

Bulk polaritons were suggested in 1958 by J. Hopfield \cite{Hopfield1958} and measured a few years later by means of non-linear optical spectroscopy \cite{Frohlich1971,Honerlage1985,Weisbuch1991,Andreani1994a}. The normal-mode coupling for polaritons is a result of momentum conservation in the exciton-photon interaction. This selection rule imposes a one-to-one coupling between exciton and photon modes having the same momentum. As a consequence, within the linear coupling regime, the situation is analogous to that of two linearly-coupled harmonic oscillators. Two normal modes at different energies, the upper and the lower polariton, are formed. Due to the very different exciton and photon energy dispersion, as a function of momentum, polariton modes display an anticrossing with a minimum energy separation that can be as large as 16 meV in bulk GaAs \cite{Andreani1994a}. Within the anticrossing region, polaritons are full admixtures of exciton and photon modes, while they have pure exciton or photon character far from it.

The concept of polaritons remained linked to the physics of bulk semiconductors for more than two decades. The progress in fabrication of epitaxial semiconductor heterostructures, particularly quantum wells (QWs) led naturally to a description of their electronic excitations in terms of the polariton concept. However, because of the mismatch in the dimensionality of excitons (2D) and photons (3D), momentum conservation applies only to the in-plane component, and one exciton mode couples to a continuum of photon modes, resulting in an irreversible radiative decay instead of a normal-mode coupling \cite{Andreani1991,Deveaud1991}. Only at larger in-plane momenta, photon modes that are evanescent in the direction orthogonal to the QW can form surface polaritons, which however result in a vanishingly small deviation from the bare exciton and photon modes \cite{Tassone1990}.

To make the leap from three to two-dimensions, polaritons had to wait until 1992, when C. Weisbuch published the first successful measurement of normal-mode coupling in a semiconductor microcavity (MC). Planar semiconductor MCs were developed in the 80's basically for producing vertical-cavity surface emitting lasers (VCSEL's). Below the lasing threshold, a VCSEL behaves as a light-emitting diode and the spontanous emission rate is determined by the details of the MC mirrors. These are stacks of layers of different semiconductor material, called distributed Bragg reflectors (DBRs), that are able to produce a high reflectivity thanks to multiple interference. One of the main problems that were being studied in connection with the MC-QW system was the change of the total spontaneous photon emission rate, starting from an initial exciton state \cite{Bjoerk1993}, that should be enhanced according to the Purcell effect \cite{Purcell1946}. The main effect of DBRs, however, is to concentrate the angle-dependent emission rate within a narrow cone around the direction normal to the sample. This led to develop mirrors of increasingly high reflectivity, resulting in a longer photon lifetime at small in-plane momenta, inside the planar structure (see e.g. the contributions by Oesterle {\em et al.} \cite{Oesterle2005} and by Joray {\em et al.} \cite{Joray2005} to the present volume). It was thanks to these advances that the exciton-photon coupling rate became faster than the damping, and the first strong-coupling sample was produced. The story behind this discovery is told by C. Weisbuch in his contribution to this volume \cite{Weisbuch2005}. 

However, normal-mode coupling was being widely investigated in another system, that of Rydberg atoms in metal resonators \cite{Berman1994,Kaluzny1983}. The differences between this system and that of polaritons are very substantial and led, in the past, to a some misunderstandings in the context of microcavity polaritons. First, an atom is a strongly non-linear object, well described in terms of a two-level system rather than a harmonic oscillator. The Rydberg transition is saturated after a single quantum of excitation has been absorbed (not accounting for spin). This intrinsic non-linearity makes the Rabi frequency, namely the rate at which the excitation is exchanged between the electromagnetic field and the atomic system, depend on the number of atoms present in the cavity. The well known Jaynes-Cummings quantum-mechanical model \cite{Jaynes1963} describes this behaviour as proportional to the square root of the number of atomic two-level systems. In this respect, the basic Rabi frequency corresponding to a single atom exchanging its energy with one cavity photon, is named vacuum-field Rabi frequency, to highlight the fact that the initial condition (and one through which the system ideally cycles during its time evolution) is that of an excited atomic transition in presence of the photon vacuum. The so called self-induced Rabi oscillations, corresponding to the regime of several atoms, had been measured already in 1983 by Kaluzny {\em et al.}. The first observation of vacuum-field Rabi oscillations for a single atom in a cavity was instead made by Thompson {\em et al.} in 1992 \cite{Thompson1992}.\footnote{Curiously enough, in the same year the semiconductor-microcavity polaritons have been also observed for the first time by Weisbuch {\em et al.}} The intrinsic non-linearity of the atom-cavity system makes the Rabi oscillations and their dependence on the number of atoms a purely quantum phenomenon, as described by the Jaynes-Cummings model. Excitons, on the other hand, are an almost non-saturable system at low density, due to the infinite spatial extension of their total wave-function in a semiconductor QW. Ideally, nonlinear effects in a QW become important when the excitons approach the saturation density \cite{Schmitt-Rink1985}. Recently, however, more subtle nonlinear effects such as the polariton parametric scattering \cite{Ciuti2001,Savvidis2000a,Stevenson2000,Ciuti2000b} have been observed in the limit of the lowest densities accessible in an optical experiment (see also the contributions by Staehli {\em et al.} \cite{Staehli2005}, by Baumberg {\em et al.} \cite{Baumberg2005}, and by W. Langbein \cite{Langbein2005} to the present volume). The second important difference between atom-cavity normal-mode coupling and bulk polaritons is the way the single-mode selection is performed. In the polariton case, the translational invariance provides the momentum selection rule that ensures single-mode coupling between photons and excitons. In the atom-cavity case, the atom is a point-like system and momentum conservation does not hold. The single-mode selection must then be engineered by photon confinement, with the requirement of a high quality factor in order for the photon escape rate to be slower than the Rabi frequency. Very recently, the same kind of vacuum-field Rabi splitting as in the atom-cavity case, was achieved by embedding one semiconductor quantum dot into a semiconductor microresonator where photons were confined in three dimensions \cite{Khitrova2006,Peter2005,Reithmaier2004,Yoshie2004} (see also the contribution by Andreani {\em et al.} \cite{Andreani2005} to the present volume, where strong cavity-quantu-dot coupling is discussed in the case of photonic-crystal microresonators). This result is the true semiconductor analogous to the atom-cavity vacuum-field Rabi splitting. 

Strong light-matter interaction and normal-mode coupling in solid-state devices are objects of increasingly intense research. The reason lies in the perspective of engineering new kinds of electronic excitations with unique quantum coherence properties, long correlation both in space and time, and robustness to environment-induced decoherence, thanks to their hybrid nature sharing the properties of light and electrons. Polaritons are the oldest manifestation of normal-mode coupling, and also one of the more promising in view of these developments. This volume collects papers by the most prominent actors in the field, providing a broad overview of the state of the art and of the directions in which polariton research is moving. The readers interested in the past evolution of the research on MC-polaritons, might look at several review articles, each focusing on different aspects of the problem \cite{Savona1999,Savona1999a,Khitrova1999,Kavokin2003d}. The present introduction aims at providing the reader with a general overview of the MC-polariton physics as it progressed during the last fifteen years. This is not, however, an exhaustive review, as certainly many important topics in polariton research are partially or not at all covered (for example the role of spin and light polarization, for which a starting point might be the contribution by Shelykh {\em et al.} \cite{Shelykh2005} to the present volume, or the recent progress in developing room-temperature polariton systems, for which the reader can refer to the contribution by Carlin {\em et al.} \cite{Carlin2005}), while many others are repeated in the individual contributions to this volume (see e.g. the contributions by R. Houdr\'e \cite{Houdre2005} and C. Weisbuch \cite{Weisbuch2005} to the present volume). It will nevertheless help the reader go through the following chapters, that are individual research reports by the most active researchers in the field.

\section{The past}

\subsection{The beginning of the microcavity polariton era}

Back in 1992, Claude Weisbuch, at the time visiting Tokyo University, proposed and realized the first semiconductor microcavity device displaying vacuum-field Rabi splitting \cite{Weisbuch1992}. The early work by Weisbuch {\em et al.} already contained all the essential features of the MC-polariton system, including the anticrossing as a function of the exciton-cavity detuning, the dependence of the Rabi splitting on the number of QWs embedded in the cavity and on the exciton and photon linewidths, and finally an analysis in terms of linear response theory. An exciting account of this discovery is given by Claude Weisbuch in this volume \cite{Weisbuch2005}, together with a short review of the early MC-polariton physics.

%\begin{figure}[ht]
%\centerline{\includegraphics[width=.47 \textwidth]{fig1.png}\includegraphics[width=.47 \textwidth]{fig1b.png}}
%\caption{Left panel: Polariton energy resonance, measured from reflectivity spectra, as a function of the exciton-cavity detuning (from Ref. \cite{Weisbuch1992}). Right panel: Polariton energy-momentum dispersion curve extracted from angle-resolved photoluminescence spectra (from Ref. \cite{Houdre1994b}).}
%\label{fig1}
%\end{figure}

The first observation of polariton vacuum-field Rabi splitting generated great excitement within a restricted community of researchers who were in quest of cavity quantum electrodynamics (CQED) effects \cite{Mabuchi2002} in semiconductor planar systems. Within a few months, further evidence of the 2-D polariton physics was provided by Houdr{\'e} {\em et al.} with the observation of the vacuum-field Rabi splitting up to room temperature \cite{Houdre1994}, and with the measurement of the energy-momentum dispersion curve by means of angle-resolved emission spectroscopy. Finally, the vacuum-field Rabi oscillations were directly time-resolved in an experiment of ultrafast photoluminescence upconversion spectroscopy by Norris {\em et al.} \cite{Norris1994} and later by Jacobson {\em et al.} \cite{Jacobson1995}.

Closely related to the observation of polaritons in MC-embedded QWs was the measurement, by Tredicucci {\em et al.}, of bulk polariton modes in a semiconductor microcavity in which the whole cavity layer provided the excitonic transition \cite{Chen1993,Tredicucci1995,Chen1995}. In this system, the cavity layer displays a series of closely spaced exciton resonances originating from the energy quantization of the exciton center-of-mass motion, confined along the growth direction within the $\lambda$-cavity slab.

Very soon after the first measurement of MC-polaritons, the very basic theoretical framework for the description of the polariton modes was developed. The appearance of a vacuum-field Rabi splitting in the polariton spectrum was understood in terms of a simple linear disperision model \cite{Houdre1993}, in full analogy to the case of atoms in an optical cavity \cite{Zhu1990}. In this model, the exciton linear response function is described in terms of a Lorentz resonance, and the overall response of the cavity--exciton system is evaluated within linear response theory of the classical electromagnetic field. In the case of atoms in a cavity, the analysis by Zhu {\em et al.} \cite{Zhu1990} suggested that the appearance of a spectral doublet in a classical linear dispersion model would somewhat ruled out the idea of vacuum-field Rabi splitting, which should be instead related to the quantum fluctuations of the field vacuum. Today we know that the linear response theory is fully equivalent to the quantum model of a Bose (non saturable) excitation coupled to the quantum field of radiation \cite{Andreani1994,Savona1995}. In this case, the vacuum field Rabi splitting is equally well described in terms of a semiclassical linear-response model. Purely quantum effects are only present if the system Hamiltonian contains many-body interaction terms (beyond the terms quadratic in the quantum fields). For the atomic case, in which the two-level system is the most appropriate description of a single atom transition, CQED effects appear as soon as the number $N$ of atoms is larger than one. This implies, in particular, an increase of the Rabi splitting -- the so called self-induced Rabi splitting -- proportional to $\sqrt{N}$ \cite{Kaluzny1983}. For polaritons, on the other hand, nonlinear effects on the Rabi splitting appear only when the exciton density approaches the saturation density, namely when $na_B^2\sim1$ \cite{Schmitt-Rink1985}, where $n$ is the exciton areal density in the QW and $a_B$ the exciton Bohr radius which, in GaAs QWs, is of the order of 10 nm. In this limit, the exciton oscillator strength and consequently the Rabi splitting vanish \cite{Jahnke1996,Khitrova1999}, contrarily to the atom-cavity case \cite{Kaluzny1983}. 

A microscopic model of the microcavity polariton modes and their energy-momentum dispersion relation was derived from the diagonalization of the linear exciton-photon coupling Hamiltonian \cite{Savona1994,Pau1995a}, within the assumption of ideal cavity mirrors, and later by including the detailed frequency response of DBRs \cite{Jorda1995,Savona1996}. These latter works presented a full three-dimensional treatment of the electromagnetic field, thus including leakage through the cavity mirrors. Within this description, polaritons arise from the linear coupling of an exciton, having a given in-plane momentum ${\bf k}_{||}$, to the continuum of photon modes having the same in-plane momentum component -- as required by the in-plane translational invariance -- and all the possible values of the remaining component $k_z$. In the case of a bare QW, this photon continuum has a smooth density of states, resulting in the intrinsic exciton radiative lifetime \cite{Andreani1991,Deveaud1991,Tassone1990}. In presence of a planar MC, the photon continuum in the $z$-direction displays a sharp peak in the density of states, corresponding to the resonant cavity mode, and the normal-mode coupling arises. The work by Savona {\em et al.} points out to the presence of {\em leaky modes} in a DBR-MC. Leaky modes arise due to resonances within the multi-layered structure formed by the cavity and the DBR layers, in which the electromagnetic field can penetrate. A particular role is played by leaky modes at frequency lower than the main cavity mode. These, due to their energy-momentum dispersion, become resonant with the exciton energy at some finite value of the in-plane momentum, typically outside the external emission cone. The leaky modes appear as sharp peaks in the emission rate of the lower polariton at large momenta \cite{Savona1996}, and in case of small exciton and photon linewidths they can even produce strong coupling with an anticrossing of normal modes. Given the two-dimensional density of states, the theory predicts that typically more than 80\% of the luminescence is emitted into the leaky modes and absorbed in the sample substrate \cite{Savona1996}.
Their presence is thus very relevant in determining the overall dynamics of the polariton photoluminescence. Strong coupling between the exciton and leaky modes was measured only very recently in a II-VI sample by Richard {\em et al.} \cite{Richard2005b}

In the same years, the equivalence between semiclassical and full quantum descriptions of MC-polaritons was finally established for the linear coupling limit \cite{Andreani1994,Savona1995,Pau1995a}. The work by Pau {\em et al.} \cite{Pau1995a} was the first to highlight the effect of an inhomogeneous distribution of exciton resonances, aimed at modeling the effect of excitons localized by interface disorder in the QW. Within the linear dispersion theory, the polariton spectrum including an inhomogeneous exciton distribution was thoroughly studied by Andreani {\em et al.} \cite{Andreani1998}. This was however only a phenomenological way of including disorder and exciton localization into the polariton problem, as the in-plane momentum conservation is lifted and a full 3-D calculation of the coupled exciton-photon modes is in principle required. This problem is briefly reviewed in the Section on structural disorder.

\subsection{Energy relaxation and polariton photoluminescence}

Most of the spectral properties of MC-polaritons are detected by angle-resolved photoluminescence, which gives access to the energy- and momentum-resolved population of the polariton states within the external emission cone. In this process, the semiconductor is optically excited at high energy, producing a population of free electron-hole pairs that eventually relax and populate the polariton states. A polariton can relax energy basically in two ways. First, through the exciton-phonon interaction, by emitting optical or acoustic phonons. Second, via many-body Coulomb interactions with other polaritons or with free carriers present in the system.

The photoluminescence dynamics is governed by the balance between energy relaxation and spontaneous emission rates. If the relaxation rates are much slower than the polariton radiative emission rates, then relaxation to the lowest states is suppressed and the spontaneous emission occurs most likely from the higher energy states. This phenomenon, called {\em relaxation bottleneck}, is expected already in the case of QW excitons \cite{Piermarocchi1996}. For polaritons, the bottleneck effect should in principle be even more effective, due to the faster radiative rates and to the steep energy-momentum dispersion in the strong-coupling region which typically suppresses inelastic scattering rates.

The bottleneck effect was observed for the first time in MCs based on II-VI materials \cite{Muller2000,Muller1999}, where the Rabi splitting is larger, and soon afterwards in III-V systems \cite{Tartakovskii2000}. In these works it was clear how the bottleneck effect was less pronounced than expected from theoretical calculations based on polariton-phonon scattering \cite{Tassone1997}, and rather strongly dependent on the excitation density. This observation suggested that many-body scattering effects take part very effectively in the energy relaxation process. Indeed, a model of the relaxation dynamics in terms of a Boltzmann equation predicts stimulated scattering to the lowest energy states in presence of polariton-polariton Coulomb scattering \cite{Tassone1999} and also if only polariton-phonon processes are taken into account \cite{Doan2002}. Stimulated polariton scattering was actually observed in experiments \cite{Dang1998,Senellart1999}. It was however clear that while stimulated scattering predicted the population buildup in the ground polariton state at high density, it could not explain the rather efficient thermalization of the polariton distribution that was typically observed at lower density.

The present understanding of this behaviour is that polariton relaxation at low to medium density is governed by at least three mechanisms. The first was proposed by Porras {\em et al.} \cite{Porras2002} and consists in an interplay between Coulomb scattering and phonon relaxation. According to this mechanism, polaritons in the strong coupling region can undergo Coulomb scattering, ending up in one polariton in a lower-energy state and the other in the exciton-like region of the dispersion, at higher energy. Within this region, which acts as a thermal reservoir, the excess energy is relaxed by acoustic phonon emission. The Porras mechanism is not driven by final-state stimulation and can therefore explain thermalization into a smooth distribution of populations. The second mechanism involved in polariton relaxation is very likely the scattering on free carriers originating from charged defects. This mechanism has been modeled within a Boltzmann formalism \cite{Malpuech2002} and evidence of it was found in experiments \cite{Tartakovskii2003,Perrin2005,Bajoni2006}. This mechanism is however strongly sample dependent, as it relies on the quality of the sample fabrication, and an accurate quantitative characterization of it is presently unavailable. As a third possible mechanism, we mention structural disorder that should give rise to polariton localization and to a lifting of the momentum conservation in the relaxation processes. Disorder in MC-polaritons is discussed in more detail below. Again, however, an accurate characterization of its influence on the relaxation process, either theoretical or experimental, is still unavailable.

\subsection{The problem of the polariton spectral linewidth}

The spectra linewidth of polaritons is determined by many factors, among which the most effective are the photon escape rate, the polariton-phonon scattering, the polariton-polariton Coulomb scattering, and the inhomogeneous spectral distribution due to structural disorder of the interfaces. The problem of polariton spectral linewidths, in the early years of MC-polaritons, has always been object of intense debate (see e.g. the account given by R. Houdr\'e in his contribution to this volume \cite{Houdre2005}).

Within the simplest possible description, consisting in a two-coupled-oscillator model, linewidths can be included as damping rates for the exciton and photon oscillators \cite{Savona1995}. Then, the frequency eigenvalues of the problem are expressed as
\begin{equation}
\omega=\frac{\omega_x+\omega_c-i(\gamma_x+\gamma_c)}{2}\pm\frac{1}{2}\sqrt{\Omega_R^2+(\omega_x-\omega_c-i(\gamma_x-\gamma_c))^2}\,,
\label{twoosc}
\end{equation}
where $\omega_x$ and $\omega_c$ are the bare exciton and cavity mode frequencies, $\Omega_R$ is the vacuum-field Rabi splitting, and $\gamma_x$ and $\gamma_c$ are, respectively, the exciton and photon damping rates. Within this purely phenomenological description, the linewidth arises as a result of a damping phenomenon and the corresponding lineshape is a Lorentzian. Eq. (\ref{twoosc}) leads, in particular, to equal polariton linewidths for zero exciton-cavity detuning $\omega_x=\omega_c$, given by the arithmetic average of the two initial damping rates. 

Although this idea was widely used in the literature, it was also strongly debated, leading to many studies of the actual polariton linewidth. In samples with bad interface quality, the actual linewidth was dominated by inhomogeneous broadening due to disorder. For high quality samples, on the other hand, experiments showed that upper and lower polariton linewidths were different even at zero detuning, with a general tendency of the lower polariton mode to have a smaller linewidth with unusually small dependency on external parameters like temperature or density \cite{Baars2000,Houdre2001,Houdre2002}.

This observation was understood as a true effect of the steep polariton dispersion. The polariton damping rate of a given state is the result of outscattering processes from that state. Scattering can occur because of phonon emission or absorption, or because of collision with another polariton. In both cases, if we assume an ideally planar system, the total energy and the total momentum are conserved in the process. Then, polaritons in the strong-coupling region of the energy momentum dispersion are characterized by a very steep dispersion curve and therefore by a very small density of final states available for an outscattering process. This mechanism of course is not included in the simple two-oscillator picture and can lead to a lower polariton mode that is extremely robust to line broadening. This was predicted by theoretical analyses in terms of the Fermi golden rule, in both cases of phonon \cite{Savona1997} and collisional \cite{Ciuti1998} broadening.

In most modern samples, because of this suppression of collisional broadening, the lower polariton linewidth is fully determined by the cavity mode linewidth.

\subsection{Influence of structural disorder}

Structural disorder in heterostructures can dramatically influence their optical response. This is the case of excitons in QWs \cite{Zimmermann2003}, where interface roughness and alloy disorder always produce localization of the center-of-mass exciton wave function over tens of nanometers. Correspondingly, an inhomogeneous distribution of the energy spectrum of the localized eigenstates arises. Exciton localization is responsible of two phenomena. The first, as already mentioned, is an inhomogeneous broadening of the optical spectrum. The second is the resonant Rayleigh scattering, namely the resonant scattering of a plane electromagnetic wave in all directions due to the breaking of the in-plane translational invariance.

In the case of MC-polaritons, the inhomogeneous broadening problem was the object of an intense and, in the light of our present understanding, overenphasized debate. All the existing works have focused on the effect of QW disorder on the exciton, while assuming an ideally planar MC. In a first work, Whittaker \cite{Whittaker1996} suggested a mechanism called {\em motional narrowing} in which the small polariton effective mass results in a long-range polariton center-of-mass wave function that averages out the exciton disorder potential, resulting in a suppression of the inhomogeneous polariton linewidth. The first theory by Whittaker was however fawled by an erroneous application of the Born approximation in perturbation theory, and Whittaker published soon a correct analysis of the problem within similar assumptions \cite{Whittaker1998}. In the meantime, Savona {\em et al.} presented a numerical result of the full diagonalization of the coupled exciton-photon Hamiltonian in presence of QW disorder \cite{Savona1997a}. Though technically correct and successful in reproducing the asymmetry in the polariton linewidth, this work included an incorrect qualitative interpretation of the numerical result in terms of interbranch scattering on disorder, that affected exclusively the upper polariton branch. In the same work, however, it was correctly pointed out that the effect of motional narrowing on the lower polariton was of the same extent as for a bare QW exciton. A more sound and simple explanation of the asymmetry funally came with the following work by Whittaker \cite{Whittaker1998}. The polariton linewidth in absorption or reflectivity experiments can be modeled within linear dispersion theory, provided one adopts the microscopically computed exciton response function in presence of disorder. This latter, because of excitonic motional narrowing, typically shows an asymmetry with a sharp low-energy tail and a more slowly decaying high-energy one \cite{Zimmermann2003}. This can easily explain the observed asymmetry and was later shown to be the correct approximation \cite{Whittaker2000} to apply when starting from the general problem \cite{Savona1997a}. This was also confirmed by an accurate simultaneous measurement of the exciton and polariton linear response \cite{Ell1998}.

Similar considerations apply to the resonant Rayleigh scattering. In this case, the MC scattering spectrum can be modeled as the bare QW exciton scattering spectrum filtered by the MC optical response \cite{Whittaker2000,Hayes1998}. 

As a general consideration, we conclude that the exciton-photon coupling does not change significantly the wave functions of localized exciton states in presence of disorder. The polariton spectral features can therefore be thought of as the result of strong coupling between the photon mode and the ensemble of all localized exciton states.

We conclude this section by pointing out that very few works have been devoted to study the effect of disorder at the interfaces of the MC. Given the strong resonant character of a planar resonator, it should be expected that thickness fluctuations of the cavity layer should give rise to an inhomogeneous photon spectrum and, to some extent, to lateral photon localization. As a result, polaritons should also be localized and inhomogeneously broadened. Evidence of the influence of photon disorder is found when comparing the measured cavity mode linewidth with the one calculated from the nominal value of the DBR reflectivity. The former is always significantly larger than the latter, suggesting an additional inhomogeneous broadening. More direct evidence of polariton localization was provided by Langbein {\em et al.} \cite{Langbein2002a}, who measured the intrinsic momentum broadening of resonant Rayleigh scattering of polaritons, deducing a polariton localization length of the order of 30 $\mu$m in a typical GaAs/AlGaAs MC. Further evidence comes from the ubiquitous cross-shaped pattern present in resonant Rayleigh scattering in momentum space \cite{Gurioli2001,Langbein2004a}. This pattern is originated from the crosshatch pattern in real space caused by misfit dislocations in the epitaxial growth \cite{Beanland1995}. Given the GaAs/AlAs lattice misfit, the typical length scale of planar regions bounded by cross hatches should be of the order of a few tens of $\mu$m, in agreement with the estimated polariton localization length in these systems. Finally, recent measurements of spatially resolved emission in presence of a large occupation of the lowest-energy polariton states in a II-VI MC \cite{Richard2005}, gave direct visual access to the localized polariton states, whose typical size in this material is of the order of a few microns.

\section{The present}

Most of the present fundamental research on MC-polaritons is focused on nonlinear optical phenomena, particularly those involving parametric polariton scattering and quantum polariton optics. The early experiments at high excitation density had prompted the experimentalists with a quite straightforward phenomenology, including an increase of the polariton linewidth and a corresponding bleaching of the Rabi splitting as a function of density \cite{Houdre1995a,Khitrova1999}. It was soon understood that the basic excitonic nonlinearities can reasonably explain these observation. In particular, the saturation of the exciton oscillator strength \cite{Schmitt-Rink1985}, well described by a Hartree-Fock model of the many-body exciton system, is responsible for the disappearance of the Rabi splitting. To the next order of scattering theory, many-body interactions also give rise to an excitation-induced dephasing, explaining the increase in linewidth. This scenario and the relevant research works are discussed in great detail in the review article by G. Khitrova {\em et al.} \cite{Khitrova1999}. In this work, the reader will also find a thorough account of the so called ``Boser controversy'', that we will briefly discuss in the following section.

\subsection{Parametric amplification and photoluminescence}

Great excitement was brought by the discovery in 2000, by the groups by J. J. Baumberg and M. S. Skolnick, of two phenomena that would mark a new era of MC-polariton physics: the polariton parametric amplification \cite{Savvidis2000a} and its spontaneous counterpart \cite{Stevenson2000}, the parametric photoluminescence. 

Polariton parametric processes bear a very close analogy with parametric downconversion of photons in nonlinear crystals, as known from the quantum optics domain \cite{Mandel1995}. In this latter case, by pumping the crystal with photons at energy $\hbar\omega$, the $\chi^{(2)}$ nonlinearity can produce two new photons of energy $\hbar\omega_1$ and $\hbar\omega_2$, whose sum is equal to the initial energy $\hbar\omega$. Depending on the geometry of the crystal, phase-matching conditions imposed by total momentum conservation have to be additionally fulfilled. In the case of polaritons, parametric processes originate from the third-order nonlinearity characterizing the polariton system. It stems from the exciton-exciton Coulomb scattering and from the density-dependent saturation of the oscillator strength, both resulting in a contribution to the Hamiltonian of fourth order in the polariton field \cite{Ciuti2003}. In the traditional setup, one laser beam is used to resonantly pump polaritons in one given mode along the lower-polariton energy-momentum dispersion curve. Given the peculiar shape of this curve -- which displays an inflection point at half energy distance from the band bottom and the bare exciton energy -- the scattering of two pump polaritons to two new polaritons, respectively at smaller and larger energy, is made possible by energy-momentum conservation. The two final states are called {\em signal} and {\em idler} polariton respectively, as in photon parametric downconversion. 

The polariton parametric amplification is the process by which, in presence of a pump, a weak probe probe beam resonant with the signal mode is amplified by stimulating the parametric process. Correspondingly, the stimulated scattering generates a polariton field in the idler mode. It is a process involving intense fields that can be fully described in terms of classical polariton field amplitudes \cite{Ciuti2003}, as in classical nonlinear optics. The first observations were obtained at the so called ``magic angle'', namely by choosing a pump momentum that results in the signal mode being at zero momentum, so that the probe beam enters the sample at normal incidence. As we will see below, however, this is not the only allowed configuration. Polariton parametric amplification was observed in samples made of different materials and can result in extremely large gain on the probe beam, reaching a few thousands at low temperature \cite{Saba2001}. It has a sharp resonant character and hence is extremely fast, allowing switching times below 1 ps even at liquid nitrogen temperature in II-VI samples. In samples with small inhomogenous broadening, it persists up to high temperature, thanks to the polariton robustness to loss mechanisms, and it can be coherently controlled \cite{Kundermann2003} (see also the contributions by Baumberg {\em et al.} \cite{Baumberg2005} and by Staehli {\em et al.} \cite{Staehli2005} to the present volume). When GaN-based polariton samples will finally be available \cite{Butte2006,Tawara2004} (see also the contribution by Carlin {\em et al.} \cite{Carlin2005} to the present volume), the amplification should persist at room temperature, thus opening the way to possible applications as fast optical switches or amplifiers.

Polariton parametric photoluminescence is a spontaneous process based on the same scattering amplitude and on the same selection rules as parametric amplification. When a polariton mode is resonantly pumped, emission is measured at the signal and idler modes even in the absence of a probe beam. The most intuitive explanation of this phenomenon is that the parametric scattering is stimulated by the vacuum field fluctuations at the signal and idler modes in the low density limit. The polariton parametric photoluminescence is an extremely complex process, one for which the theoretical modeling was essential for understanding the rich phenomenology it displays. The most considerable contribution to the theory of parametric processes and photoluminescence in particular was given by C. Ciuti and is reviewed in Ref. \cite{Ciuti2003}. In the basic theory, the pump polariton field is assumed to be a purely classical field, driven by the resonant laser beam. By further neglecting polariton scattering terms that do not involve the pump mode, the resulting Hamiltonian is quadratic in the signal and idler polariton operators $\hat{p}_{{\bf k}}$ and $\hat{p}_{{\bf k}_i}$, however containing anomalous terms that do not conserve the particle number. Here ${\bf k}$ is the signal momentum, ${\bf k}_i=2{\bf k}_p-{\bf k}$ is the idler momentum and ${\bf k}_p$ is the pump momentum fixed by the external pump angle. Then, the dynamical equations for the signal and idler polariton populations $N_{{\bf k}}=\langle\hat{p}_{{\bf k}}^\dagger\hat{p}_{{\bf k}}\rangle$ and $N_{{\bf k}_i}=\langle\hat{p}_{{\bf k}_i}^\dagger\hat{p}_{{\bf k}_i}\rangle$ are linearly coupled to an analogous equation for the anomalous quantum correlation $A_{{\bf k}}=\langle\hat{p}_{{\bf k}}^\dagger\hat{p}_{{\bf k}_s}^\dagger\rangle$. Here we report these equations for clarity (restricted to the lower polariton branch),
\begin{eqnarray}
\frac{dN_{{\bf k}}(t)}{dt}&=&-\frac{2\gamma_{{\bf k}}}{\hbar}N_{{\bf k}}+\frac{2}{\hbar}\mbox{Im}\left[gP^{2}_{{\bf k}_p}\mbox{e}^{-2i\omega_pt}A_{{\bf k}}+\cal{L}_{{\bf k},N}\right]\label{normal}\\
i\hbar\frac{dA_{{\bf k}}(t)}{dt}&=&\left[-E_{{\bf k}}-E_{{\bf k}_i}-i(\gamma_{{\bf k}}+\gamma_{{\bf k}_i})\right]A_{{\bf k}}-gP^{*2}_{{\bf k}_p}\mbox{e}^{2i\omega_pt}(1+N_{{\bf k}}+N_{{\bf k}_i})+\cal{L}_{{\bf k},A}\label{anomalous}
\end{eqnarray}
where $E_{{\bf k}}$ is the polariton dispersion, $\gamma_{{\bf k}}$ the parametrized polariton damping rate (accounting for radiative and nonradiative mechanisms), $\omega_p$ is the pump frequency, $P_{{\bf k}_p}$ is the classical pump polariton field, and $g$ is the nonlinear coupling strength assumed $k$-independent for simplicity. The quantities $\cal{L}_{{\bf k},(A,N)}$ derive from quantum Langevin random forces that are needed in the formalism, if the damping terms are included. From these equations, it appears clearly that the polariton populations are driven by the pump only through the anomalous correlation $A_{{\bf k}}$. This proves that parametric photoluminescence is a purely quantum process and cannot be interpreted in terms of a Boltzmann equation involving solely population variables, as was done by some authors \cite{Erland2001}. A very detailed experimental characterization by W. Langbein \cite{Langbein2004} on a high-quality sample, in particular, showed that the increase of signal and idler populations in a pulsed experiment is not instantaneous, following the increase of the anomalous term (see also the contribution by W. Langbein to this volume \cite{Langbein2005}). Many features can be inferred from Eqs. (\ref{normal}) and (\ref{anomalous}). For a fixed pump momentum ${\bf k}_p$, the parametric process is allowed for a set of values of the signal and idler momentum, forming an ``eight''-shaped curve in momentum coordinates. This pattern was measured for the first time very recently by W. Langbein \cite{Langbein2004}. This ``eight''-shaped pattern was never observed before, as a sample of extremely good quality was required. In particular, the polariton energy-momentum dispersion must be well defined not only in the strong coupling but also in the exciton-like region. For QWs with some amount of disorder, the inhomogeneous distribution both in energy and momentum, in the exciton-like region of the dispersion, will completely wash out the selection rule for the parametric process. The sample adopted by W. Langbein was designed with state-of-the-art exciton inhomogeneous broadening and is, up to now, the only sample on which the parametric momentum pattern has been measured. If two pump beams are used to pump different polariton modes, than mixed parametric processes arise in which two polaritons, each from a different pump, scatter to a signal and a idler mode. The energy-momentum conservation in this case can give rise to a rich variety of shapes of the momentum pattern, that were measured on the same sample \cite{Savasta2005} (see also W. Langbein's contribution to this volume \cite{Langbein2005}).

If Fourier-transformed to the frequency domain, Eqs. (\ref{normal}) and (\ref{anomalous}) predict the resonant frequencies of the parametric photoluminescence. It turns out \cite{Ciuti2003} that for large pump intensity these resonances are shifted with respect to the bare polariton dispersion, and can display bifurcations or a change in the sign of the first derivative of the dispersion curve. Evidence of this behaviour has been found in experiments \cite{Savvidis2001}. 

In Eq. (\ref{anomalous}), the term $(1+N_{{\bf k}}+N_{{\bf k}_i})$ shows that for $N>1$ the parametric photoluminescence displays a stimulated behaviour, which was experimentally characterized with high accuracy \cite{Langbein2004}. Far above the stimulation threshold, Eqs. (\ref{normal}) and (\ref{anomalous}) no longer hold and will diverge for a finite value of the pump polariton amplitude. In more realistic terms, the stimulated behaviour will deplete the pump polariton mode at a rate faster than the input rate provided by the external pump beam. In this case the equations can be generalized by introducing a third equation for the classical pump field, with the external laser field amplitude as the pump parameter. If this is done, then the pump mode in the model is depleted before reaching the critical value for which the equations diverge. A very interesting issue however arises in connection with the stimulated behaviour. It is expected that above threshold a spontaneous symmetry breaking occurs and the polariton field at the signal and idler modes acquires a classical complex amplitude $\langle\hat{p}_{{\bf k}}\rangle\ne0$. This is called {\em polariton parametric oscillation}. It can be interpreted as a poor men's Bose-Einstein condensation (BEC), with the classical field amplitude at the signal mode being the order parameter for the phase transition \cite{Baumberg2000}. Indeed, below threshold the phases of the signal and idler quantum fields are purely fluctuating, while only their {\em relative} phase is fixed by the link to the classical pump amplitude. C. Ciuti was the first to suggest the possible occurrence of parametric oscillations by noticing that a singular solution with a non-zero classical field is admitted by the parametric photoluminescence equations, if the pump polariton amplitude takes exactly the critical value. Since however the parametric photoluminescence equations (\ref{normal}) and (\ref{anomalous}) assume purely quantum fluctuating fields and diverge for the pump amplitude approaching this critical value, they cannot describe the way this spontaneous symmetry breaking occurs for increasing pump intensity. Later, Savona {\em et al.} \cite{Savona2005a} extended the parametric photoluminescence theory in the spirit of the Hartree-Fock-Bogolubov approximation \cite{Shi1998}, commonly adopted to model BEC of an interacting gas. This model was limited by the restriction to three polariton modes, but allowed for the first time a description of the transition to the parametric oscillation regime. A quantum montecarlo study closely followed \cite{Carusotto2005}, generalizing this result to all polariton modes. The polariton field amplitudes in the parametric oscillation are expected to behave in all respects as a quantum fluid, in particular featuring superfluidity \cite{Ciuti2005a,Baas2006}.

If parametric oscillations at the signal and idler modes occur, the corresponding classical field amplitudes can in turn act as pump fields for parametric photoluminescence involving other polariton modes. This gives rise to off-branch resonances and multiple scattering \cite{Ciuti2003,Whittaker2005}. This behaviour was observed in experiments \cite{Savvidis2001}, and represents a proof -- though indirect -- that parametric oscillations with spontaneous symmetry breaking do take place. Further indirect evidence was found in the observation of optical bistability in parametric oscillations \cite{Baas2004a,Baas2004b}.

Finally, we point out to a recent experiment \cite{Diederichs2006} where parametric oscillation was observed in a very special setup of three vertically stacked semiconductor MCs in the weak coupling regime.

\subsection{Quantum correlation and non-classical properties}

Probably the most appealing feature of polariton parametric photoluminescence is that signal-idler polariton pairs are produced in non-classical states with quantum correlations. These have their origin in the expression of the parametric Hamiltonian, that reads
\begin{equation}
\hat{H}=\sum_{\bf k}E_{\bf k}\hat{p}_{{\bf k}}^\dagger\hat{p}_{{\bf k}}+\sum_{{\bf k},{\bf k}^\prime,{\bf k}_s,{\bf k}_i}\left[G_{{\bf k},{\bf k}^\prime}\hat{p}_{{\bf k}_s}^\dagger\hat{p}_{{\bf k}_i}^\dagger+\mbox{H.c.}\right]\delta_{{\bf k}_s+{\bf k}_i,{\bf k}+{\bf k}^\prime}\,,
\label{parham}
\end{equation}
where
\begin{equation}
G_{{\bf k},{\bf k}^\prime}=gP_{{\bf k}}P_{{\bf k}^\prime}(\delta_{{\bf k},{\bf k}_{p1}}+\delta_{{\bf k},{\bf k}_{p2}})(\delta_{{\bf k}^\prime,{\bf k}_{p1}}+\delta_{{\bf k}^\prime,{\bf k}_{p2}})
\end{equation}
is a parametric coupling term generalized to the case of two pump fields at momenta ${\bf k}_{p1}$ and ${\bf k}_{p2}$. The signal and idler momenta are now denoted ${\bf k}_{s}$ and ${\bf k}_{i}$. If one evaluates the time-evolution of the system, starting from the vacuum polariton state as the initial state, then in the limit of low pump amplitude (or equivalently in the limit of short time) the quantum state of the system will be
\begin{equation}
|\psi(t)\rangle=\alpha(t)|0\rangle+\sum_{{\bf k}_s}\beta_{{\bf k}_s}(t)|1_{{\bf k}_s}1_{{\bf k}_i}\rangle+\ldots\,,
\label{quantumstate}
\end{equation}
where $\beta\ll\alpha$. In this state, signal and idler polaritons are pair-correlated, namely the joint probability of having a signal polariton {\em and} and idler polariton is equal to the probability of just having one of them \cite{Karr2004,Savasta2005}. Again, this scenario is closely related to what happens in quantum optics, where the analogous property of parametric photon downconversion has been used to produce pair correlated and entangled photons, even in the limit of intense laser beams \cite{Lamas-Linares2001}.

Very recently there were several proposals of taking advantage of this feature of polariton parametric photoluminescence for observing non-classical states of polaritons. One of the first observations of this kind was the amplitude squeezing of light emitted from parametric oscillation, by Karr {\em et al.} \cite{Karr2004a}. Indeed the parametric Hamiltonian, with the assumption of a classical pump field, is identical to that of a nonlinear Kerr medium \cite{Mandel1995} and is therefore able to produce a quantum state for which the Heisenberg uncertainty of one quadrature operator is smaller (by only 5\% in this experiment) than that of its canonical conjugate, always fulfilling the uncertainty principle. The experiment by Karr {\em et al.} was performed in the special setup in which pump, signal and idler polariton modes coincide at zero momentum -- a singular point in momentum space for which parametric scattering is always allowed. This result was confirmed by a theoretical analysis \cite{Schwendimann2003}. It should be pointed out that polariton squeezing was already predicted, although in the limit of negligible many-body interactions, long before by Hradil {\em et al.} \cite{Hradil1993}. Intrinsic polariton squeezing can in principle occur because of the anti-resonant terms in the minimal coupling Hamiltonian between radiation and matter in the Coulomb gauge. However, the theory predicts an extremely small amount of squeezing in this case, that cannot be detected in realistic situations on GaAs-based materials. A full account of polariton squeezing, both instrinsic and in parametric oscillations, can be found in the contribution by A. Quattropani and P. Schwendimann to the present volume \cite{Quattropani2005}.

Another expression of non-classical physics that can arise from parametric photoluminescence is polariton entanglement. By inspection of the quantum state (\ref{quantumstate}), we see that it is a linear combination of the vacuum field and of an entangled state all possible signal-idler pair-states allowed by the conservation rules. Hence, momentum entanglement is a natural outcome of polariton parametric photoluminescence. In a similar way, spin entanglement is also possible. In fact, the polariton-polariton scattering is spin conserving, and the exciton-radiation selection rules are such that spin-up (-down) polaritons couple to clockwise (counterclockwise) circularly polarized light \cite{Ciuti1998a}. By using a linearly polarized pump beam then, simple linear superposition principle shows that the signal-idler polariton pair will be generated in a linear combination of the two possible spin states. This kind of polarization entanglement would require photon-coincidence detection to be measured and is still awaiting experimental confirmation. Ciuti recently proposed a special kind of momentum entanglement called branch entanglement \cite{Ciuti2004}. In this case the entanglement arises with respect to the polariton branch index, and can be obtained by setting the pump resonant with the upper polariton branch. However, as Ciuti pointed out, a pump beam resonant with the upper polariton will generate a large amount of outscattering to the exciton-like part of the lower polariton branch at large momenta, implying a fast loss rate that might make an experimental verification challenging. The only evidence of polariton momentum entanglement was obtained by W. Langbein in an experiment where a two-pump setup was used \cite{Savasta2005}, following a theoretical prediction by Savasta {\em et al.} (an account of this experimental result can also be found in the contribution by W. Langbein to this volume \cite{Langbein2005}). The scope of this experiment was to prove Bohr's quantum complementarity principle on polaritons. To this purpose, the two idler polaritons produced by the two pump beams were made interfere on a detector. In the low intensity limit, below the parametric oscillation threshold, at most one signal-idler polariton pair at a given time is present. Then, the same situation as in Young's two-slit experiment exists. The two idler-polariton paths will produce quantum interference if and only if the two paths are indistinguishable. This happens only if the two corresponding signal modes coincide. Differently, it would be possible -- by independently detecting one of the two signal polaritons -- to gather ``which-way" information on the two idler channels and, quantum interference could not arise. A similar experiment was successfully carried out in the case of parametrically downcoverted photons by Mandel {\em et al.} \cite{Mandel1999}. The experimental proof of this mutual exclusion between ``which-way'' information and quantum interference was an indirect evidence that, in the case of two distinguishable signal polaritons, the polariton pair is produced in a momentum-entangled state. The possibility of fabricating polariton quantum boxes (see next section) could make it possible, by the same principle, to produce spatially entangled pairs of confined polaritons over distances of several microns.

It should be pointed out that a polariton is a many-body electronic excitation of a semiconductor device, existing in a densely packed medium and therefore in principle subject to strong decoherence induced by coupling to the environment. The ability to generate nonclassical states of electronic excitations holds great promise for future implementations of quantum information technology in semiconductor devices. Polaritons, due to their hybrid nature and robustness to decoherence, might be the best candidate for achieving this purpose.

To conclude this section, we point out to a general difficulty in measuring the idler-polariton photoluminescence in parametric experiments in the standard single-pump configuration. This is due basically to three factors, all related to the fact that the idler mode is located in the exciton-like region of the polariton dispersion. First, the idler is almost fully exciton-like, with a small photon component, thus suppressing the coupling to the external electromagnetic field and the photoluminescence intensity. Second, outscattering mechanisms are more favoured by the flat dispersion curve, and the damping rate of the idler polariton mode is consequently faster than that of the signal mode. Third, disorder induced inhomogeneous broadening, both in energy and momentum, is only effective in the exciton-like part of the polariton dispersion, thus lifting the momentum selection rule and partially washing out the sharp resonance of the idler photoluminescence. This feature is very general, as already remarked. More emphasis should therefore be devoted to experiments in the double pump geometry or using quantum-confined polariton levels in quantum boxes (see below), where these limitations are not present.

\section{The future}

\subsection{Polariton quantum collective phenomena}

The perspective of observing polariton quantum collective phenomena is the driving force of present and future research on MC-polaritons. By quantum collective phenomenon we mean a phase transition resulting in a spontaneous symmetry breaking with formation of a quantum mechanical order parameter. Such a situation is very appealing because it would imply the existence of a macroscopic quantum state of many polaritons, displaying properties such as superfluidity, Josephson oscillations in confined geometry, robustness to decoherence. In addition to the fundamental interest of this novel quantum collective phenomenon in solids, it would also bear a great potential for applications in devices exploiting the quantum phase, like quantum information processing.

In the previous section we have already mentioned the occurrence of spontaneous symmetry breaking in the transition from the regime of polariton parametric photoluminescence to parametric oscillation. If, on one side, this is at present the only polariton quantum collective phenomenon backed up by convincing experimental evidence, its behaviour is not governed by the physics of phase transitions in thermal equilibrium. In particular, temperature and chemical potential cannot be defined for this system, although in the limit of the generalized Gross-Pitaevskii, analogous to the zero-temperature Bogolubov model of BEC, the pump energy can be interpreted as a pinned chemical potential, in the sense of the ground-state energy of the quantum fluid. Further, the signal and idler polariton gases below threshold, though being each an incoherent gas of quasiparticles, still have quantum pair-correlations. Nevertheless, the parametric oscillator is expected to display quantum fluid behaviour and superfluidity, as was described by Ciuti and Carusotto \cite{Carusotto2005} (see also their contribution to this volume \cite{Ciuti2005b}).

An intense research activity is being instead devoted to the quest for polariton BEC \cite{Deng2002,Deng2003,Doan2005,Eastham2006,Eastham2001,Kavokin2003c,Keeling2004,Laussy2004,Laussy2006a,Malpuech2003a,Marchetti2006,Marchetti2004,Porras2002,Richard2005,Richard2005a,Savona2005,Snoke2002,Zamfirescu2002} (see also the contributions by Baumberg {\em et al.} \cite{Baumberg2005}, Ciuti {\em et al.} \cite{Ciuti2005b}, Savona {\em et al.} \cite{Savona2005}, and Shelykh {\em et al.} \cite{Shelykh2005} to the present volume). As in the case of excitons, for which BEC has been sought for about fourty years (see e.g. Ref. \cite{Griffin1995} and the special issue of Solid State Communications vol. 134 (2005)), the idea was stimulated by the quasi-bosonic behaviour of polaritons in the limit of low density, and by their extremely small effective mass. This topic is widely reviewed later in this volume by Savona and Sarchi \cite{Savona2005}, and we would not indulge in a repetition of the same concepts. Let us just highlight a few important points concerning the two-dimensional character, the non-equilibrium nature and what would be conclusive experimental evidence in a spectroscopy experiment.

The Hohenberg-Mermin-Wagner theorem \cite{Hohenber1967} states that conventional BEC with the formation of off-diagonal long-range order cannot take place in two dimensions. The reason is that thermal fluctuations of the quantum phase would dominate at any finite temperature, fully destroying the condensate. No long-range order can then arise. However, it can be shown that for an interacting gas, still a superfluid behaviour can be achieved below a critical temperature $T_c$. In this case, a different kind of phase transition is instead expected: the Berezinskii-Kosterlitz-Thouless transition \cite{Pitaevskii2003}. This is a topological phase transition in which, above a critical temperature $T_{BKT}$, vortices can spontaneously form (thanks to the fact that a vortex in two dimensions costs a finite amount of energy), destroying superfluidity. The critical temperature $T_{BKT}$ is always lower than the superfluid  critical temperature $T_c$ of the spatially homogeneous gas, and is therefore the phase transition that should be observed in two-dimensions. The Berezinskii-Kosterlitz-Thouless transition has been discussed in the context of MC-polaritons \cite{Kavokin2003c}. This interpretation should however be revisited in the light of disorder and polariton localization, always occurring in MCs. The Berezinskii-Kosterlitz-Thouless is a topological transition and makes sense only in a perfetly homogeneous medium. Polariton localization, as mentioned above, occurs within the range of a few to a few tens of microns \cite{Langbein2002a,Richard2005}. A rigorous proof holds \cite{Lauwers2003} that whenever in a two-dimensional system a finite energy gap separates the ground single-particle state from the excited states, then thermal fluctuations no longer dominate and conventional BEC with a macroscopic occupation of the ground state is recovered. Such a gap can typically arise in confined systems (see discussion below), or even in presence of disorder because locally, within the area interested by optical excitation, polariton localization will give rise to a discrete energy spectrum for the lowest-lying states. Even in the limit of a system of infinite extension, it can be rigorously proved \cite{Lenoble2004} that an ideal Bose gas can undergo conventional BEC, thanks to the Lifshitz tail in the lowest part of the density of states, causing condensation to occur in some localized state at finite values of temperature and density. Both confinement and polariton localization implie a finite spatial extension of the condensate.

The deviations from thermal distribution of polaritons, in connection to the relaxation bottleneck effect, have already been discussed. A debate is currently ongoing, on whether polariton BEC can be understood in terms of standard equilibrium thermodynamics or if deviations from equilibrium occur. A conclusive answer to this question is not yet available. However, if on one hand it is clear that at sufficiently high excitation density the relaxation mechansims taking place through many-body interactions will favour thermalization within the polariton branches \cite{Porras2002,Tartakovskii2003}, this does not exclude that a large non-equilibrium population of hot excitons or electron-hole pairs at large energy and momentum exists. This hot population, while not affecting the condensate thermalization, could nevertheless modify the polariton spectrum via many-body scattering and by saturating the exciton oscillator strength \cite{Schmitt-Rink1985}. Another deviation from thermal equilibrium are quantum fluctuations of the condensate, that are expected to occur even at zero temperature. These typically tend to deplete the condensate by occupying single-particle excited states \cite{Leggett2001,Pitaevskii2003,Dalfovo1999}. In the case of strong many-body interactions, this condensate depletion can be very effective and produce deviations from the standard Bose-Einstein distribution of polariton populations.

Finally, let us briefly discuss what would be the ``smoking gun'' of polariton BEC in an optical experiment. In the case of diluted alkali atoms, the striking observation was a peaked distribution in momentum space of the atomic population \cite{Anderson1995}. This was very remarkable as, for a common atomic gas, classical statistical mechanics and the Gibbs principle make this particular configuration in the system phase space extremely unlikely. There are however other unique features characterizing a Bose condensate. First, the energy-momentum dispersion of collective excitations, displaying the typical Bogolubov linear behaviour at small momenta \cite{Steinhauer2002}. Second, the off-diagonal long-range order, expressed in the spatial quantum correlations, that are nonzero up to infinite distance \cite{Bloch2000a}. These correlations are today considered as the true distinctive feature of an interacting condensate, according to the Penrose-Onsager criterion \cite{Penrose1956}. Third, related to the spatial coherence, there is the ability of two condensates to produce quantum interference when overlapping \cite{Andrews1997}. All these features were clearly observed in the case of diluted atoms, but were simply considered as further confirmations of the occurrence of BEC. For polaritons, observing a macroscopic occupation of the ground state might be not enough. The reason is twofold. First, a coherent pump beam is present in the system at high energy, and might induce a coherent excitation of the ground state by some unexpected many-body scattering effect (we have just seen how this can happen in parametric oscillations). Second, a semiconductor at high enough excitation density will start lasing, with the result of emitting light in a single mode of the momentum space. This ambiguity, in particular, was at the basis of the ``boser'' controversy, that followed an observation of strong single-mode emission under high density excitation \cite{Pau1996} (for a detailed account, see Khitrova {\em et al.} \cite{Khitrova1999}). The observation was later explained in terms of a bleaching of the exciton-photon coupling and a subsequent lasing effect \cite{Kira1997}, and the same authors withdrew their initial interpretation \cite{Cao1997b}. Credit must be given, however, to the work by Imamoglu and Yamamoto \cite{Imamoglu1996}, who have for the first time suggested the possibility of MC-polariton BEC. Later, a similar strong nonlinear emission at zero momentum under high-energy nonresonant excitation was reported by several authors, in more controlled experiments where simultaneous evidence of polaritons in the strong coupling regime was provided \cite{Senellart1999,Dang1998,Boeuf2000,Bleuse1998}. Although very promising, these observations are not sufficient to claim BEC, as explained. More recently, polariton BEC was claimed by Deng {\em et al.} \cite{Deng2003,Deng2002}, again on the basis of a momentum redistribution of polaritons, but with simultaneous measurement of the spatial intensity profile and of the second-order time coherence of the emitted light. Yet, these observations are not conclusive, as none of the distinguishing features mentioned above has been evidenced, while the measured second-order time coherence does not display a sharp transition at the supposed condensation threshold, casting therefore doubts on the validity of this claim. The most recent experiments are focusing on coherence properties in real and momentum space \cite{Richard2005a,Richard2005}, and leave hope that a conclusive observation of polariton BEC is not very far ahead.

\subsection{Engineering quantum confined polariton nanodevices}

The history of semiconductor physics has been marked by a continuous progress towards reducing the dimensionality of semiconductor structures, from bulk semiconductors, to quantum wells, quantum wires, eventually producing a rich variety of zero-dimensional nanostructures. These advances were driven by the idea of confining electrons in order to obtain a quantized spectrum with discrete energy levels that would display novel optical and transport properties to be exploited in nanodevice technology.

Polaritons in bulk materials were first described by Hopfield in 1958 \cite{Hopfield1958}, but it took until 1992 \cite{Weisbuch1992} before the idea of two-dimensional polaritons in MCs was conceived and realised! The reason is probably to be attributed to the need for high-quality semiconductor optical resonators that were made available only with the advances in epitaxial growth. The two further steps, towards one-dimensional and zero-dimensional polariton systems, are the object of very recent research.

Zero-dimensional confinement of polaritons would be extremely attractive both for fundamental studies and in view of applications. On the fundamental side, we have already remarked that a discrete polariton spectrum could be favorable to BEC and to parametric polariton photoluminescence or oscillations with an even balance between signal and idler. Pairs of polariton quantum boxes with tunnel coupling might display Josephson oscillations as soon as a quantum fluid -- either a Bose condensate or a parametric oscillation -- is formed. They would also allow the design of several configurations in which polariton spatial entanglement can be achieved. It should be also pointed out that polaritons, due to their very steep dispersion, would display quantum confinement and a discrete energy spectrum already when confined over a few microns. This would make fabrication, optical and electronic addressing of such a microdevice an extremely simple task, if compared to semiconductor quantum dots \cite{Bimberg1999}. Among the possible applications, one could imagine devices for single photon emission, as well as for the generation of entangled pairs of electronic excitations or the storage of quantum information, that could be employed in quantum information technology.

The studies on confined polaritons are very recent and mostly based on the idea of etching a {\em micropillar} out of a planar semiconductor MC. This approach is justified by the fact that the micropillar fabrication technology can benefit of the progress in vertical-cavity surface-emitting lasers (VCSELs). The experimental efforts were also motivated by the quest for strong coupling and the formation of normal modes between a zero-dimensional cavity mode and an interband transition in a quantum dot \cite{Reithmaier2004}. The transition from two-dimensional polaritons to one and zero dimension was studied for the first time by Dasbach {\em et al.} \cite{Dasbach2003,Dasbach2001,Dasbach2002a} and by Obert {\em et al.} \cite{Obert2004,Obert2004a}. In the experiments by Dasbach {\em et al.}, both one-dimensional wires and pillars were studied by photoluminescence spectroscopy. The one-dimensional result clearly displayed polariton dispersion features, with the simultaneous presence of upper and lower polariton branches \cite{Dasbach2003,Dasbach2002a}. The single-micropillar spectrum clearly showed discrete energy modes from which it was possible to observe parametric photoluminescence \cite{Dasbach2003,Dasbach2001}. In this case, however, a clear signature of simultaneous upper and lower polariton modes was not reported. Evidence of upper and lower polariton is essential to claim the formation of strong coupling and exciton-photon normal modes, as parametric scattering can be equally obtained in the weak coupling regime \cite{Diederichs2006}. The reason why polariton features were not observed in these early studies is perhaps to be attributed to the low quality factor of the resonator, introduced by etching the whole body of the microcavity. This technology has however progressed very quickly in the last years and presently micropillars can be fabricated with quality factors approaching $10^5$ \cite{Loffler2005}. Although these new microresonators would probably display striking polariton features, no samples with embedded quantum wells have been produced so far.

It is interesting to notice, however, that the common scheme used in today's microcavity samples, consisting in producing two-dimensional confinement of both excitons and photons, is redundant, as polaritons already exist in the bulk material and their energy-momentum dispersion is dominated by the steep photon mode dispersion. Ideally, by reducing the dimensionality of a bulk polariton system, confined polaritons could be obtained. Indeed, this idea is the starting point of the realisation of bulk MC-polaritons, by Tredicucci {\em et al.} \cite{Tredicucci1995}, where only the photon mode is confined to two dimensions, the excitonic transition being provided by the bulk cavity layer. An analogous idea to produce polariton quantum boxes has recently been introduced by El. Da\"if {\em et al.} \cite{Daif2006}. Starting from a planar MC, photon modes can be confined by engraving a very shallow pattern on top of the cavity layer and growing the top mirror afterwards. Thicker regions on the MC plane will have a photon-mode resonance at a frequency lower than in thinner regions. Thus, by patterning e.g. a circular region thicker than the surrounding, this will act as a lateral potential well for the photon modes, eventually producing confinement. If the planar MC has one or more embedded quantum wells and displays polariton modes, then confined polaritons will arise by virtue of this mechanism. A very shallow pattern is enough to produce confinement of a few polariton modes. As an example, the sample in Ref. \cite{Daif2006} has circular mesa patterns of 6 nm height and three different diameters: 3.6, 9 and 19 $\mu$m. The 6 nm height corresponds in a GaAs MC to a variation of the photon mode energy of about 9 meV. As a result, the smallest diameter mesa clearly shows three confined modes of the lower polariton, and three corresponding confined modes of the upper polariton. The energy quantization amounts to about 1 meV in this sample, and it is smaller (with correspondingly more extended modes) for the two larger mesas. The interest in this kind of samples lies in the presence of both confined and extended modes, due to the finite height of the energy barrier in the confining potential. The photoluminescence measurements clearly show both lower and upper confined polaritons, displaying an anticrossing behaviour as the exciton-cavity detuning is varied. The spectral lines are very narrow, suggesting that the interfaces are very flat within the mesa region. These polariton quantum boxes can thus reduce the polariton linewidth to the minimum value determined by the cavity mode lifetime. Thanks to the ease in fabrication and optical addressing, this new technique is very promising for the realisation of systems of coupled polariton quantum boxes that are an ideal system for the observation of polariton quantum fluids and nonclassical states.

\section{Conclusions}

We have reviewed the most important developments in the field of MC-polaritons during the last fifteen years. Started as a straightforward extension of three-dimensional polariton physics, this area of research has now become extremely active after the recent developments in polariton parametric effects, the possibility of Bose-Einstein condensation, and the evolution towards polariton nanodevices. It involves an increasingly large number of research groups in the world, who are currently exploring new fronteers along the directions of nanostructure engineering and the use of new materials. The feeling within the community -- beautifully expressed by the various contributions to the present volume -- is that the time is not far when polariton devices will become a perfect workbench for studying fundamental quantum physics, and a prominent technology for modern quantum optoelectronics.

%\bibliographystyle{plain}
%\bibliography{MC2}

\end{document}